\documentclass[a4paper,10pt,preprint,twocolumn,3p]{elsarticle} 
\usepackage[justification=Justified]{caption}
\usepackage{threeparttable}
\usepackage{graphicx}
\usepackage{dcolumn}
\usepackage{multirow}
\usepackage[dvipsnames]{xcolor}
\usepackage{bm}
\usepackage[dvipsnames]{xcolor}
\usepackage[colorlinks=true]{hyperref}
\usepackage{natbib}
\usepackage{gensymb}
\usepackage{lineno}
\usepackage{mathtools}
\usepackage{bm,amssymb}
\usepackage{subcaption}
\usepackage{booktabs}
\usepackage{color} 
\usepackage{natbib}
\usepackage{physics}
\usepackage{siunitx}
\usepackage[dvipsnames]{xcolor}

\usepackage{comment}

\journal{Physics Letters B}

\biboptions{numbers,sort&compress}
\begin{document}
\onecolumn
\begin{frontmatter}
\title{Two-neutrino $\beta\beta$ decay to excited states at next-to-leading order}

\author[1,2,3]{Daniel Castillo\corref{cor1}}
\ead{daniel.castillo@fqa.ub.edu}

\author[1,2,3]{Dorian Frycz\corref{cor1}}
\ead{dorianfrycz@fqa.ub.edu}

\author[3]{Beatriz Benavente}
\ead{beabdl2000@gmail.com}

\author[1,2]{Javier Men\'{e}ndez}

\ead{menendez@fqa.ub.edu}

\address[1]{%
 Departament de F\'isica Qu\`antica i Astrof\'isica , Universitat de Barcelona, c. Martí i Franquès, 1, 08028 Barcelona, Spain
}
\address[2]{
Institut de Ci\`encies del Cosmos, Universitat de Barcelona, c. Martí i Franquès, 1, 08028 Barcelona, Spain
}%

\address[3]{Facultat de Física, Universitat de Barcelona, c. Martí i Franquès, 1, 08028 Barcelona, Spain}

\cortext[cor1]{Corresponding author.}

\date{\today}

\begin{abstract}
We study two-neutrino double-beta decay ($2\nu\beta\beta$) into first-excited $0^+_2$ states of nuclei used in $\beta\beta$ decay experiments, including $^{76}$Ge, $^{82}$Se, $^{130}$Te, and $^{136}$Xe. We calculate the corresponding nuclear matrix elements (NMEs) within the nuclear shell model, using various Hamiltonians that describe well the spectroscopy of the initial and final nuclei. 
We evaluate the next-to-leading order (NLO) long-range NMEs recently introduced within chiral effective field theory, keeping three terms in the expansion of the energy denominator. In most cases, NLO contributions to the half-life are below 5\%, but they can {significantly} increase due to cancellations in the leading-order Gamow-Teller NME. 
A detailed analysis in terms of nuclear deformation, including triaxiality, indicates that larger deformation differences between the initial and final states generally lead to smaller NMEs, but the seniority structure of the states also plays a relevant role. The lower range of our predicted half-lives, with uncertainties dominated by the nuclear Hamiltonian used, are slightly longer than the current experimental limit in $^{76}$Ge and {consistent with} the very recent half-life indication in $^{82}$Se.

\end{abstract}
\end{frontmatter}

\section{ Introduction}\label{Introduction}
In two-neutrino double-beta ($2\nu\beta\beta$) decay two neutrons convert into two protons. This process is accompanied by the emission of two electrons  and two antineutrinos  simultaneously~\cite{Saakyan:2013yna,Engel17}:
\begin{equation}
\begin{split}
    &(A, Z)\xrightarrow{}(A, Z+2)+2e^-+2\bar{\nu}_e\,,
    \end{split}
    \label{eq:2vbb}
\end{equation}
and represents the slowest decay ever measured, with half-lives that can exceed $10^{21}$ years~\cite{Barabash23, Barabash20}. 

$2\nu\beta\beta$ decay is closely related to the neutrinoless $\beta\beta$ decay, where no antineutrinos are emitted. The latter is a hypothetical decay not allowed within the standard model of particle physics, as it violates lepton number conservation. Its observation would be groundbreaking, confirming that neutrinos are their own antiparticles and providing insights into the matter-antimatter asymmetry in the universe \cite{Agostini23,Gomez-Cadenas:2023vca}. The connection between the two $\beta\beta$ modes appears not only because both involve the same initial and final states, but in addition various studies have found that the nuclear matrix elements (NMEs) for both decays are correlated~\cite{Jokiniemi:2022ayc, Horoi:2022ley,Horoi:2023uah}. A detailed understanding of $2\nu\beta\beta$ decay is thus key to improve predictions for the neutrinoless $\beta\beta$ mode.

$2\nu\beta\beta$ decays have been measured in several nuclei, mainly for ground-state-to-ground-state transitions ($0^+_{\rm gs}\xrightarrow{}0^+_{\rm gs}$)~\cite{PhysRevD.93.112008, GERDA23, CUPID23, NEMO-Se, PhysRevC.98.024617, CUPID-Mo23, NEMO-319, Barabash18, LZ:2024wvs, PhysRevC.106.024328, CUORE:2025xue, PhysRevC.105.055501, PandaX-4TGS, KamLAND-Zen19,  PhysRevD.94.072003}. However, theoretical calculations are challenged by the evaluation of a second-order weak process that depends on both initial and final states as well as the intermediate odd-odd nucleus. In fact, most many-body methods, such as the nuclear shell model (NSM)~\cite{Caurier:1990dc,Poves:1995rg,Horoi:2007xe,CAURIER201262,Horoi:2013jx}, the quasiparticle random-phase approximation (QRPA) method~\cite{EJIRI20191,Hinohara:2022uip}, or the interacting boson model (IBM)~\cite{PhysRevC.91.034304}, overestimate $2\nu\beta\beta$ decay NMEs. This is because missing nuclear correlations and the lack of two-nucleon currents, as demonstrated in Refs.~\cite{Gysbers2019,Stroberg:2021guc} for Gamow-Teller (GT) $\beta$ decays. Notable exceptions are an effective field theory for $\beta$ decays (EFT$_\beta$)~ \cite{CoelloPerez:2024dzt, Jokiniemi:2022yfr, PhysRevC.106.034309, PhysRevC.98.045501}, with couplings fitted to GT data, and calculations with renormalized operators using coupled cluster theory~\cite{PhysRevLett.126.182502} and the NSM~\cite{Coraggio:2022vgy, Coraggio:2018tuo}, that can also include two-nucleon currents~\cite{Coraggio24}. 

Nonetheless, the overestimation, or need for {\it quenching}, of the NMEs is systematic within the same mass region~\cite{Wildenthal83, Chou93, Martinez96}, and common to GT and $2\nu\beta\beta$ decays{. It} can be accounted for phenomenologically. Indeed, this approach allowed the theoretical prediction by the NSM of the $^{48}$Ca $2\nu\beta\beta$ decay~\cite{Caurier:1990dc,Poves:1995rg} and double-electron capture in $^{124}$Xe~\cite{CoelloPerez:2018ghg} before their measurement---the latter was also predicted by the EFT$_\beta$ and QRPA~\cite{XENON:2019dti}.   

In this work, we 
calculate within the NSM the $2\nu\beta\beta$ half-lives of ground-state-to-excited-state ($0^+_{\rm gs}\xrightarrow[]{}0^+_{2}$) decays in $^{48}$Ca, $^{76}$Ge, $^{82}$Se, $^{124}$Sn, $^{130}$Te, and $^{136}$Xe. Previous NSM works have studied $^{48}$Ca~\cite{PhysRevC.87.014320}, $^{76}$Ge~\cite{gerda_collaboration_2_2015}, $^{124}$Sn~\cite{horoi_shell_2016}, and $^{136}$Xe~\cite{Jokiniemi:2022yfr}, besides $^{48}$Ca, $^{76}$Ge, and $^{82}$Se with a renormalized operator and two-nucleon currents~\cite{Coraggio24}. $2\nu\beta\beta$ decays to excited states have also been explored with the EFT$_\beta$~\cite{Jokiniemi:2022yfr, PhysRevC.98.045501}, QRPA~\cite{PhysRevC.91.054309,Jokiniemi:2022yfr} and IBM~\cite{PhysRevC.91.034304,Jokiniemi:2022yfr}.
In order to obtain more reliable results, we follow an improved formalism expanding the lepton energies~\cite{Simkovic18}. Further, in the spirit of Ref.~\cite{Castillo:2024jfj}, which explored next-to-next-to-leading order neutrinoless $\beta\beta$ decay NMEs, we evaluate for the first time in these decays next-to-leading order (NLO) $2\nu\beta\beta$ terms derived within chiral effective field theory ($\chi$EFT)~\cite{morabit20242nubetabetaspectrumchiraleffective}. 

$2\nu\beta\beta$ $0^+_{\rm gs}\xrightarrow[]{}0^+_{2}$ decays have been measured in $^{100}$Mo \cite{Augier23}, $^{150}$Nd \cite{Polishcuk21}, and there is a very recent experimental indication for $^{82}$Se \cite{Barabash:2025bxa}. Lower half-life limits have been set for instance in $^{48}$Ca~\cite{Bakalyarov:2002wu}, $^{76}$Ge \cite{Arnquist24}, $^{82}$Se \cite{Arnold20}, $^{116}$Cd \cite{Barabash18}, $^{124}$Sn \cite{DAWSON2008167}, $^{130}$Te \cite{Adams21}, and $^{136}$Xe \cite{PANDAX-4T, AlKharusi_2023, ASAKURA2016171}. To better compare with experimental data, we consider various nuclear Hamiltonians that describe well the structure of both initial and final states, {and two different GT operators. First, the bare GT one, with NMEs that we need to quench phenomenologically~\cite{Caurier:1990dc,Poves:1995rg,CoelloPerez:2018ghg}. In addition, an effective GT operator renormalized within the NSM valence space~\cite{Coraggio:2018tuo}. Since this effective operator is based on a Hamiltonian evolved to high cutoff, a mild contribution from two-nucleon currents is anticipated~\cite{CoelloPerez:2018ghg,Gysbers2019}. We account for this possible effect by phenomenologically demanding that $2\nu\beta\beta$ half-lives to ground states are reproduced. While we do not include two-nucleon currents explicitly, we expect that their impact---smaller for the effective GT operator---is within our theoretical uncertainties. 
}

In addition, we explore the impact of nuclear deformation, an aspect extensively discussed for neutrinoless $\beta\beta$ decay~\cite{Menendez:2008jf,Rodriguez:2010mn,Fang:2011da,Mustonen:2013zu,Song:2014vra,Yao:2016oxk,Yao:2019rck,Wang:2021shq,Tsunoda:2023fqw,Jiao:2023poq,Wang:2023hkc} but less so for the $2\nu\beta\beta$ mode~\cite{Alvarez-Rodriguez:2004ueg,Yousef:2008jw,Nitescu:2024ruw,Hinohara:2022uip}. In particular, we analyze the shape invariants of the initial and final nuclei, taking into account the impact of triaxiality and configuration mixing on the $2\nu\beta\beta$ NMEs.

\section{Two-neutrino $\beta\beta$ decay}

At leading order, the $2\nu\beta\beta$ half-life is,
\begin{equation}\label{eq:T_LO}
    \left(T^{2\nu}_{1/2}\right)^{-1}=g^4_A G^{2\nu}_0\left(M^{(-1)}_{\rm GT}\right)^2\,,
\end{equation}
with $g_A=1.27$ the axial coupling, $G^{2\nu}_0$ the phase-space factor (PSF), and $M^{(-1)}_{\rm GT}$ the leading-order NME. This expression can be improved in two ways.

First, we follow Ref.~\cite{Simkovic18} and Taylor expand the lepton energies, introducing higher-order terms that couple additional PSFs and subleading GT NMEs. The latter are given by \cite{Simkovic18},
\begin{align}
     &M^{(-2m-1)}_{\rm GT}=m_e(2m_e)^{2m} \nonumber \\
     \times\sum_n&\frac{\langle 0^+_f||\sum_j{{\text{GT}^k_j}\,\tau^-_j}\color{black}||1^+_n\rangle\langle 1^+_n||\sum_i  {{\text{GT}_i^k}}\,\tau^-_i\color{black}||0^+_i\rangle}{(E_n-(E_i+E_f)/2)^{2m+1}},
\end{align}
where $|0^+_i\rangle$ and $|0^+_f\rangle$ denote the angular momentum-parity $J^\pi=0^+$ initial and final states, with energies $E_i$ and $E_f$, respectively, and $|1^+_n\rangle$ represents each available state of the intermediate  odd-odd nucleus, with energy $E_n$.  {We use two different GT operators: the bare GT operator${, \text{GT}^{\rm bare}_j}=\boldsymbol{\sigma}_j$,} where $\boldsymbol{\sigma}$ is the spin operator{, and the effective GT operator, ${\text{GT}^{\rm eff}_j}$, taken from Ref.~\cite{Coraggio:2018tuo}}. $\tau^-$ is the isospin ladder operator and  $m_e$ the electron mass. We keep additional NMEs up to $m\leq 2$, $M^{(-3)}_{\rm GT}$ and $M^{(-5)}_{\rm GT}$, which are sufficient to converge our results to the percent level. These NMEs have been explored for $0^+_{\rm gs}\xrightarrow{}0^+_{\rm gs}$ decays in $^{136}$Xe~\cite{KamLAND-Zen19}, $^{100}$Mo~\cite{CUPID-Mo23}, and $^{130}$Te~\cite{CUORE:2025xue} in studies of the $2\nu\beta\beta$ spectrum.

\begin{table*}[t]
    \centering
    \caption{PSFs and $\mathcal{Q}$~\cite{Ovidiu21, NNDC} for all $2\nu\beta\beta$ $0^+_{\rm gs}\xrightarrow{}0^+_{2}$  decays studied in this work, and the $0^+_{\rm gs}\xrightarrow{}0^+_{\rm gs}$ decay of $^{124}$Sn. Columns $3$ to $6$ list the PSFs arising in the lepton-energy expansion, while the final column 
    corresponds to the weak-magnetism PSF. }
    \begin{tabular}{ccccccc}
    \hline\hline
        Decay & $\mathcal{Q}$ (MeV) & $G^{2\nu}_0$ (yr$^{-1}$) & $G^{2\nu}_2$ (yr$^{-1}$)& $G^{2\nu}_4$ (yr$^{-1}$)& $G^{2\nu}_{22}$ (yr$^{-1}$)& $G^{2\nu}_M$ (yr$^{-1}$)\\
        \hline
        $^{48}$Ca $(0^+_{\rm gs}\xrightarrow[]{}0^+_{\rm 2})$ &$1.2709$ &$3.523\cdot 10^{-22}$  &$2.911\cdot 10^{-23}$  & $2.986\cdot 10^{-24}$ &$7.250\cdot 10^{-25}$ &$2.015\cdot 10^{-24}$ \\
        
        $^{76}$Ge $(0^+_{\rm gs}\xrightarrow[]{}0^+_{\rm 2})$ &$0.9168$ &$6.949\cdot 10^{-23}$ &$3.070\cdot 10^{-24}$ &$1.671\cdot 10^{-25}$ &$4.273\cdot 10^{-26}$ &$3.299\cdot 10^{-25}$\\
        
         $^{82}$Se $(0^+_{\rm gs}\xrightarrow[]{}0^+_{\rm 2})$ &$1.5102$ &$4.555\cdot 10^{-21}$ &$5.370\cdot 10^{-22}$ &$7.839\cdot 10^{-23}$ &$1.908\cdot 10^{-23}$  &$2.828\cdot 10^{-23}$\\
        
         $^{124}$Sn $(0^+_{\rm gs}\xrightarrow[]{}0^+_{\rm gs})$  & $2.2927$  &$5.628\cdot 10^{-19}$  &$1.509\cdot 10^{-19}$ &$5.028\cdot 10^{-20}$ &$1.194\cdot 10^{-20}$ &$4.529\cdot 10^{-21}$\\

        $^{124}$Sn $(0^+_{\rm gs}\xrightarrow[]{}0^+_{\rm 2})$ & $0.6354$ & $2.117\cdot 10^{-23}$  & $4.536\cdot 10^{-25}$ & $1.194\cdot 10^{-26}$ & $3.228\cdot 10^{-27}$ & $8.541\cdot 10^{-26}$\\
     
        $^{130}$Te $(0^+_{\rm gs}\xrightarrow[]{}0^+_{\rm 2})$ &$0.7340$&$7.592\cdot 10^{-23}$ &$2.168\cdot 10^{-24}$ &$7.607\cdot 10^{-26}$  &$2.034\cdot 10^{-26}$ &$3.231\cdot 10^{-25}$ \\
        
         $^{136}$Xe $(0^+_{\rm gs}\xrightarrow[]{}0^+_{\rm 2})$ &$0.8788$&$3.621\cdot 10^{-22}$ &$1.479\cdot 10^{-23}$  &$7.429\cdot 10^{-25}$ &$1.956\cdot 10^{-25}$ &$1.660\cdot10^{-24}$ \\
         \hline\hline
    \end{tabular}
    \label{tab:PSF}
\end{table*}

Second, we include NLO $\chi$EFT contributions to $2\nu\beta\beta$ decay very recently introduced by Ref.~\cite{morabit20242nubetabetaspectrumchiraleffective}. This involves weak magnetism along with one-pion-exchange diagrams, leading to~\cite{morabit20242nubetabetaspectrumchiraleffective}, 
\begin{align}\label{eq:Total_half_life}
    &(T^{2\nu}_{1/2})^{-1}=g^4_A\left(M_{\rm GT}^{(-1)}\right)^2\Delta_0\left[G^{2\nu}_{0}+\xi_{31}\frac{\Delta_2}{\Delta_0}G^{2\nu}_2\right. \nonumber \\
    &\left. +G^{2\nu}_4\left(\xi_{51}\frac{\Delta_2}{\Delta_0}+\frac{1}{3}\xi_{31}^2\right)+G^{2\nu}_{22}\frac{1}{3}\xi_{31}^2+\frac{G^{2\nu}_{M}}{\Delta_0} \right]\,.
\end{align}
The PSFs from the lepton expansion, $G^{2\nu}_i$, are available for $0^+_{\rm gs}\xrightarrow{}0^+_{\rm gs}$ decays~\cite{Simkovic18, Ovidiu21}, along with the magnetic PSF, $G_{M}^{2\nu}$~\cite{morabit20242nubetabetaspectrumchiraleffective}. For $0^+_{\rm gs}\xrightarrow{}0^+_{2}$ decays, we compute all PSFs solving numerically the Dirac equation with the RADIAL code \cite{salvat2019radial}, following the procedure of Ref.~\cite{Ovidiu21}. Table~\ref{tab:PSF} lists our PSFs and the $Q$-value ($\mathcal{Q}$) for all decays studied in this work. 

The subleading NMEs enter through the ratios $\xi_{31}={M^{(-3)}_{\rm GT}}/{M^{(-1)}_{\rm GT}}$ and $\xi_{51}={M^{(-5)}_{\rm GT}}/{M^{(-1)}_{\rm GT}}$. In addition, $\Delta_0$ and $\Delta_2$, defined as,
\begin{align}\label{eq:Delta}
            &\Delta_0=1+\frac{4}{g^2_A}\frac{3\epsilon_{\rm GT}-\epsilon_{\rm F}}{M^{(-1)}_{\rm GT}}-\frac{2g_M}{3m_Ng_A}(\mathcal{Q}+2m_e)\,, \nonumber \\
        &\Delta_2=1+\frac{2}{g^2_A}\frac{3\epsilon_{\rm GT}-\epsilon_{\rm F}}{M^{(-1)}_{\rm GT}}\,,
\end{align}
 combine NLO one-pion-exchange NMEs,
\begin{align}
\frac{3\epsilon_{\rm GT}-\epsilon_{\rm F}}{g^2_A}&=\frac{m_e }{8\pi F^2_\pi R_A}\Bigl(M^{AA}_{\rm GT}(m_\pi)-2M^{AP}_{\rm GT}(m_\pi) \nonumber \\
        & +4M^{PP}_{\rm GT}(m_\pi)-\frac{g^2_V}{g^2_A}M_{\rm F}(m_\pi)
        \Bigr)\,,    
\end{align}
because we do not account for NLO short-range NMEs proportional to currently-unknown couplings, or small tensor contributions~\cite{morabit20242nubetabetaspectrumchiraleffective}. Here $g_M=4.7$, $g_V=1$ are the isovector magnetic and vector couplings, $m_N=939$ MeV, $m_\pi=138.4$ MeV denote the nucleon and pion mass, $F_{\pi}=92.28$ MeV is the pion decay constant, and $R_A=1.2 A^{1/3}\text{ fm}$ is the nuclear radius, with $A$ the nucleon number. The one-pion-exchange NMEs are, 
\begin{align}\label{eq:NME_NLO_corrections}
    M^{X}_{k}(m_\pi)&=\frac{2R_A}{\pi}\langle 0^+_f||\sum_{ij}\tau_i^-\tau_j^-\mathcal{O}_k\int^\infty_0 dp \frac{p^2}{p^2+m_\pi^2} \nonumber \\
    &\times h^{X}_k(p^2)
    j_0(p\,r_{ij})||0^+_i\rangle\,,
\end{align}
where the spin structure is given by $\mathcal{O}_{\rm GT}=\boldsymbol{\sigma}_i\boldsymbol{\sigma}_j$ or $\mathcal{O}_{\rm F}=\mathbf{I}_{ij}$, $j_0$ is the spherical Bessel function, $r_{ij}$ the distance between the two decaying neutrons, and $p$ the momentum transfer. 
We add short-range correlations missing in the NSM many-body states~\cite{Weiss:2021rig} via Argonne and CD-Bonn parameterizations~\cite{Simkovic:2009pp}, but their impact in our results is minor. Finally, just as in neutrinoless $\beta\beta$ decay~\cite{Agostini23}, 
we have,
\begin{align}
        h_{\rm F}(p^2)=&h^{AA}_{\rm GT}(p^2)=1,\quad h^{AP}_{\rm GT}(p^2)=-\frac{2}{3}\frac{p^2}{p^2+m_\pi^2}\,, \nonumber \\
        & h^{PP}_{\rm GT}(p^2)=\frac{1}{3}\frac{p^4}{(p^2+m^2_\pi)^2}\,.
\end{align}

\section{Nuclear Shell Model}
\label{sec:NSM}

We calculate all many-body states within the NSM~\cite{Shellmodel,brown1988status,otsuka2020evolution}. For each nucleus, we consider standard shell-model Hamiltonians tailored for a given isospin-symmetric valence space that well reproduce the low-lying spectroscopy of the initial and final nuclei. For $^{48}$Ca, we employ the interactions KB3G \cite{Poves:2000nw} and GXPF1A \cite{Honma2002} in the $pf$ shell $\{0f_{7/2},1p_{3/2},0f_{5/2},1p_{1/2}\}$ orbitals. For $^{76}$Ge and $^{82}$Se, we employ the JJ4BB \cite{JJ4B}, GCN2850 \cite{Caurier:2007wq}, JUN45 \cite{JUN45}, and RG \cite{prolate} Hamiltonians in the orbitals $\{1p_{3/2},0f_{5/2},1p_{1/2},0g_{9/2}\}$. Finally, for $^{124}$Sn, $^{130}$Te, and $^{136}$Xe, we use GCN5082 \cite{Caurier:2010az} and QX \cite{Qi2012} in the orbitals $\{0g_{7/2},1d_{5/2},2s_{1/2},1d_{3/2},0h_{11/2}\}$. Throughout our work, we have employed the ANTOINE and NATHAN codes~\cite{ANTOINE,Shellmodel}.

As indicated by previous studies, the NSM systematically overestimates the NMEs for GT~\cite{Martinez96,Chou93,Wildenthal83}  and $2\nu\beta\beta$ decays~\cite{Caurier:1990dc,Poves:1995rg,Horoi:2007xe,CAURIER201262,Horoi:2013jx} {when using the bare GT operator.} We correct for this by reducing the value of our $2\nu\beta\beta$ $0^+_{\rm gs}\rightarrow 0^+_2$ NMEs by the same quenching factor, $q_{\beta\beta}$, that our NSM results require to reproduce the experimental $2\nu\beta\beta$ $0^+_{\rm gs}\rightarrow 0^+_\text{gs}$ half-life in the same nucleus~\cite{Barabash23, Barabash20}. For $^{124}$Sn, where $2\nu\beta\beta$ decay has not been measured, we use the $q_{\beta\beta}$ values obtained for $^{130}$Te, $^{136}$Xe, and the double-electron capture in $^{124}$Xe~\cite{CoelloPerez:2018ghg,Nitescu:2024ppf}. When available, we also consider the factor, $q_{\beta}$, needed to reproduce a set of GT decays in the same mass region~\cite{CAURIER201262}. 

{Alternatively, the effective GT operator from Ref.~\cite{Coraggio:2018tuo} captures additional many-body correlations in the valence space. In addition to the results obtained directly with ${\text{GT}^{\rm eff}_j}$, we consider the impact of possible deficiencies, including missing two-nucleon currents, by phenomenologically introducing  the $q_{\beta\beta}$ corresponding to the effective GT operator.} Our NSM NMEs are finally,
\begin{equation}
    M^{(-2m-1)}_{{\rm GT}, \text{NSM}}=q^2_{i}\, M^{(-2m-1)}_{\rm GT}\,,
\end{equation}
{with $q_i$ representing either $q_{\beta\beta}$ or $q_{\beta}$ for bare-GT NMEs, and unquenched $q=1$ or $q_{\beta\beta}$ for effective-GT NMEs.} 

Nuclear deformation, evidenced by rotational bands connected by strong electromagnetic transitions, is very common across the nuclear chart. Thus, accurately characterizing deformation is crucial for understanding nuclear structure~\cite{Garrett:2021kfb,heyde2011shape}. 
Shape invariants offer a robust way to do so~\cite{Kumar:1972zza,Cline:1986ik}. They provide a model-independent framework to quantify deformation parameters and their fluctuations, via rotationally-invariant products of quadrupole $Q_{2}$ operators,
\begin{align}
    Q^n=[[Q_2\times Q_2...\times Q_2]_2\times Q_2]_0\,,
\end{align}
where subscripts indicate the rank of the (coupled) operators, resulting in a scalar quantity.
For instance, the absolute deformation parameter $\beta_2$ is related to $Q^2$, and its softness to $Q^4$. Likewise, the triaxiality parameter $\gamma$ requires the knowledge of $Q^3$ and its softness is related both to $Q^5$ and $Q^6$. We apply the method explained in Ref. \cite{poves2020limits} to calculate within the NSM $\beta_2$ and $\gamma$ along with their fluctuations.

\section{Results and Discussions}
\begin{figure}[t]
 \centering
      \includegraphics[width=0.99\linewidth]{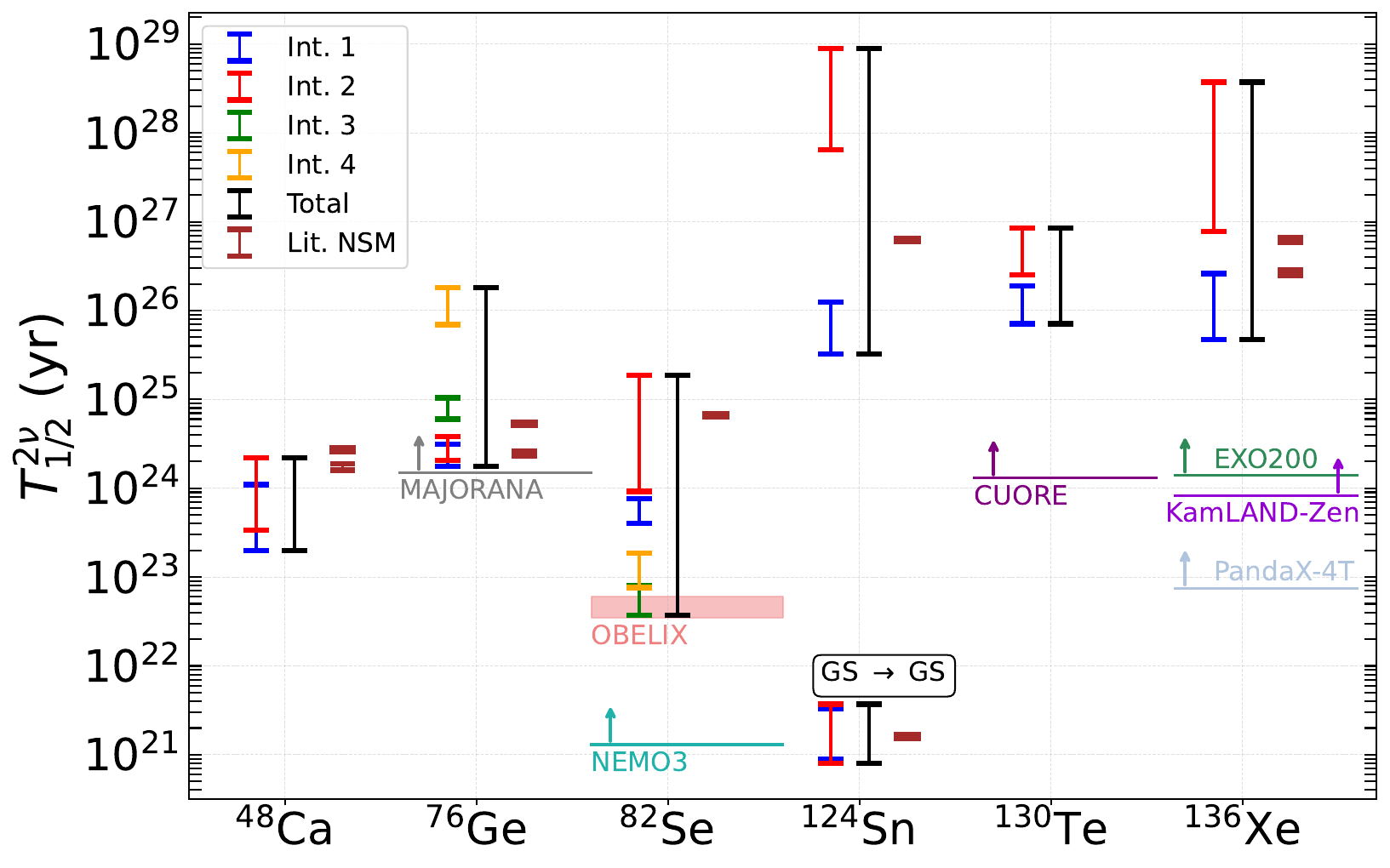}
    \caption{$2\nu\beta\beta$ half-lives for all $0^+_{\rm gs}\xrightarrow[]{}0^+_{2}$ decays studied in this work, besides $0^+_{\rm gs}\xrightarrow[]{}0^+_{\rm gs}$ for $^{124}$Sn. We use various nuclear interactions (indicated by bars with different colors) in each mass region. \textcolor{black}{The associated uncertainties span the results obtained with bare and effective GT operators and the range of quenching factors $q_{\beta}-q_{\beta\beta}$ ($1-q_{\beta\beta}$) for the bare (effective) operator (see text).} The total uncertainty (black bars) encompasses all our predictions. They are compared to 90\% confidence level experimental limits~\cite{Arnquist24, Arnold20, Adams21, DAWSON2008167, AlKharusi_2023, ASAKURA2016171, PANDAX-4T,Bakalyarov:2002wu} (horizontal bars with arrows), the $^{82}$Se measurement indication at 1$\sigma$~\cite{Barabash:2025bxa} (wide band), and other NSM predictions~\cite{Coraggio24, PhysRevC.87.014320,gerda_collaboration_2_2015,horoi_shell_2016,Jokiniemi:2022yfr} (brown bars). Limits for $^{48}$Ca, $^{124}$Sn are below the scale.}
    \label{Results_half_lives}
\end{figure}

\begin{table*}[t]
    \centering
    \caption{ $2\nu\beta\beta$ half-lives (third \textcolor{black}{and fourth} columns) for all decays studied in this work (first column), obtained for different shell-model interactions (second column) and range of quenching factors, \textcolor{black}{$q_{\beta}-q_{\beta\beta}$ for GT$^{\rm bare}$ and $1-q_{\beta\beta}$ for GT$^{\rm eff}$, given in \ref{AnnexB}}. $\varepsilon_{\rm Taylor}$ (fifth \textcolor{black}{and sixth} columns) indicates the relative half-life contribution from the expansion of the lepton energies, and $\varepsilon_{\rm NLO}$ (seventh \textcolor{black}{and eighth} columns) the one of NLO terms.}
\begin{tabular}{
    c            
    l            
    S[table-format=3.2(3)]  
    S[table-format=3.2(3)]  
    S[table-format=2.1]     
    S[table-format=2.1]     
    @{\hskip 0.5cm}S[table-format=2.1(2)]  
    @{\hskip 0.7cm}S[table-format=2.1(2)]  
}
\hline\hline
Decay & Interaction  &  \multicolumn{2}{c}{$T^{2\nu}_{1/2}$ ($10^{23}$yr)} & \multicolumn{2}{c}{$\varepsilon_{\rm Taylor}$ (\%)} & \multicolumn{2}{c}{$\varepsilon_{\rm NLO}$ (\%)}  \\
\hline\noalign{\vskip 3pt}
 & & \multicolumn{1}{c}{\text{GT$^{\rm bare}$}} & \multicolumn{1}{c}{\text{GT$^{\rm eff}$}} & \text{GT$ ^{\rm bare}$} & \text{GT$^{\rm eff}$} & \text{GT$^{\rm bare}$} & \text{GT$^{\rm eff}$}  \\
\hline \noalign{\vskip 3pt}
\multirow{1}{*}{$^{48}$Ca } & KB3G  & 8.4(14) &  6.5(45) & 0.8 & 0.8 & 1.4(2) &  1.1(5) \\
$(0^+_{\rm gs}\xrightarrow[]{}0^+_{\rm 2})$& GXPF1A   & 16.6(46) & 12.7(93) & 1.2 & 1.2 & 1.9(4)& 1.6(8)  \\[3pt]
\hline \noalign{\vskip 2pt}
\multirow{3}{*}{$^{76}$Ge }
& GCN2850    & 19.9(18) & 24.5(70) & 1.0 & 1.2 & 1.3(1) &  1.4(3)  \\
 & JUN45      & 22.7(20) &  30.5(77) & 1.2 & 1.2 & 1.4(2) & 1.5(4)  \\
$(0^+_{\rm gs}\xrightarrow[]{}0^+_{\rm 2})$& JJ4BB          & 64.0(1) & 82(22) & 1.8 & 2.3 & 2.0(1) & 2.3(4) \\
& RG      & 716(22)  & 1670(130) & -1.0 & -2.7 & 4.1(28)& 6.2(42)  \\[3pt]
\hline \noalign{\vskip 2pt}
\multirow{3}{*}{$^{82}$Se }
& GCN2850    & 6.0(16)&  4.3(3)   & 11& 12  & 0.07(4)& 0.06(4)  \\
& JUN45      & 11.8(26)&  180(8)    & 14& 9.3  & 0.5(4)& -2.6(12)   \\
$(0^+_{\rm gs}\xrightarrow[]{}0^+_{\rm 2})$ & JJ4BB           & 0.744(1)& 0.59(21)  & 7.6& 8.3 & 0.10(2)& 0.07(5)   \\
& RG      & 1.659(1)&  1.31(55)    & 3.5& 3.4 & 0.4(1)& 0.3(2)  \\[3pt]
\hline \noalign{\vskip 2pt}
\multirow{1}{*}{$^{124}$Sn }
& GCN5082    & 0.019(10) & 0.021(12) & 4.2& 3.7 & 5.9(20)& 6.1(22) \\
$(0^+_{\rm gs}\xrightarrow[]{}0^+_{\rm gs})$ & QX     & 0.0132(52)& 0.024(13) & 3.0& 2.7 & 4.4(17)& 5.8(27)\\[3pt]
\hline \noalign{\vskip 2pt}
\multirow{1}{*}{$^{124}$Sn }
& GCN5082    & 760(410) &  780(460) & 1.2 & 1.2 & 2.0(7)& 2.0(7)  \\
$(0^+_{\rm gs}\xrightarrow[]{}0^+_{\rm 2})$ & QX     & ${6.9(31)\cdot 10^{4}}$& ${52(37)\cdot 10^4}$    & 4.1& 1.9 & -19(10)& -64(46)     \\[3pt]
\hline \noalign{\vskip 2pt}
\multirow{1}{*}{$^{130}$Te }
& GCN5082    & 1350(530)&  1040(330)   & 0.9 & 1.2  & 0.8(2)& 0.7(2)    \\
$(0^+_{\rm gs}\xrightarrow[]{}0^+_{\rm 2})$ & QX          & 5370(20)& 5520(3000)      & 3.0& 1.9 & 1.1(6)&  1.2(9)  \\[3pt]
\hline \noalign{\vskip 2pt}
\multirow{1}{*}{$^{136}$Xe }
& GCN5082   & 1700(910)& 1110(640)    & -6.0& -3.9 & 7.2(23)& 6.0(21)   \\
$(0^+_{\rm gs}\xrightarrow[]{}0^+_{\rm 2})$ & QX   &  7960(210)& ${22(15)\cdot 10^{4}}$     & 7.8 & 0.3 & -18.9(35) &  -142(93)  \\[3pt]
\hline\hline
\end{tabular} \label{tab:results_tab}
\end{table*}

\subsection{$2\nu\beta\beta$ decay half-lives}\label{sec:Total_half_lives}
Figure~\ref{Results_half_lives} and Table~\ref{tab:results_tab} summarize our main results for the $2\nu\beta\beta$ half-lives to excited $0_2^+$ states. Figure~\ref{Results_half_lives} shows our results in bars colored according to the shell-model interaction used (second column in Table~\ref{tab:results_tab}): KB3G, GCN2850, and GCN5082 in blue; GXPF1A, JUN45, and QX in red; JJ4BB in green and RG in yellow. The error bar for each calculation arises from {considering both bare and effective GT operators, and their corresponding quenching values, given in \ref{AnnexB}---for the renormalized operator this includes not having any quenching. The impact of short-range correlations is minor.}
When combined, the overall uncertainty (black bars) stems {mainly} from the choice of nuclear Hamiltonian. For comparison, Fig.~\ref{Results_half_lives} also shows current experimental limits~\cite{Arnquist24, Arnold20, Adams21, DAWSON2008167, AlKharusi_2023, ASAKURA2016171, PANDAX-4T,Bakalyarov:2002wu} (horizontal bars with arrows), the $^{82}$Se measurement indication~\cite{Barabash:2025bxa} (wide band), and other NSM predictions~\cite{Coraggio24,PhysRevC.87.014320,gerda_collaboration_2_2015,horoi_shell_2016,Jokiniemi:2022yfr} (brown bars). 
The latter are consistent with our results, except in $^{48}$Ca when using renormalized operators and currents~\cite{Coraggio24}. 
\ref{Annex} collects the results of other many-body methods~\cite{PhysRevC.87.014320,PhysRevC.91.034304,PhysRevC.98.045501,gerda_collaboration_2_2015,horoi_shell_2016,PhysRevC.91.054309}.

{In general, the bare and renormalized operators yield rather similar results with moderate error bars. The latter is because $q_{\beta\beta}$ and $q_\beta$ are similar for GT$^\text{bare}$, and also because the GT$^\text{eff}$ reproduces reasonably well the measured $2\nu\beta\beta$ half-lives to ground states.} {Thus, in most} cases the uncertainty of our predictions is dominated by the nuclear Hamiltonian. {The exception is} $^{48}$Ca, {where} shell-model Hamiltonians are better refined after decades of systematic studies~\cite{Shellmodel}. 

{We only observe noticeable deviations between the results obtained with bare and effective GT operators in few cases: $^{82}$Se with the JUN45 interaction, and  $^{124}$Sn and $^{136}$Xe with QX. In these decays, $T^{2\nu}_{1/2}$ for GT$^\text{eff}$ are longer by approximately one order of magnitude due to large cancellations in the running sum of the leading GT NMEs. Thus, our total uncertainties for the last two nuclei span over three orders of magnitude.}
{Nonetheless, for these transitions}, the better reproduction of the low-lying spectrum of the intermediate nucleus, $^{136}$Cs~\cite{Rebeiro:2023kvs}, and the successful prediction of the $^{124}$Xe half-life~\cite{CoelloPerez:2018ghg} indicate a preference for the GCN5082 Hamiltonian (blue bars). {For this interaction, longer half-lives correspond to $q_{\beta\beta}$ (bare operator) and the unquenched renormalized operator. These calculations also better describe} the low-lying GT strength in $^{136}$Xe~\cite{CAURIER201262}. Nonetheless, {the} NSM predictions remain two orders of magnitude longer than current experimental limits, suggesting that near-term detection in these nuclei may be unlikely.

The comparison is more interesting for $^{76}$Ge and $^{82}$Se, since the lower range of the NSM half-life approaches the current experimental limit for $^{76}$Ge and {is consistent with} the $^{82}$Se measurement claim. Nevertheless, the uncertainty due to the Hamiltonian is substantial, spanning over {two orders} of magnitude.

\begin{figure}[t]
 \centering
    \includegraphics[width=0.99\linewidth]{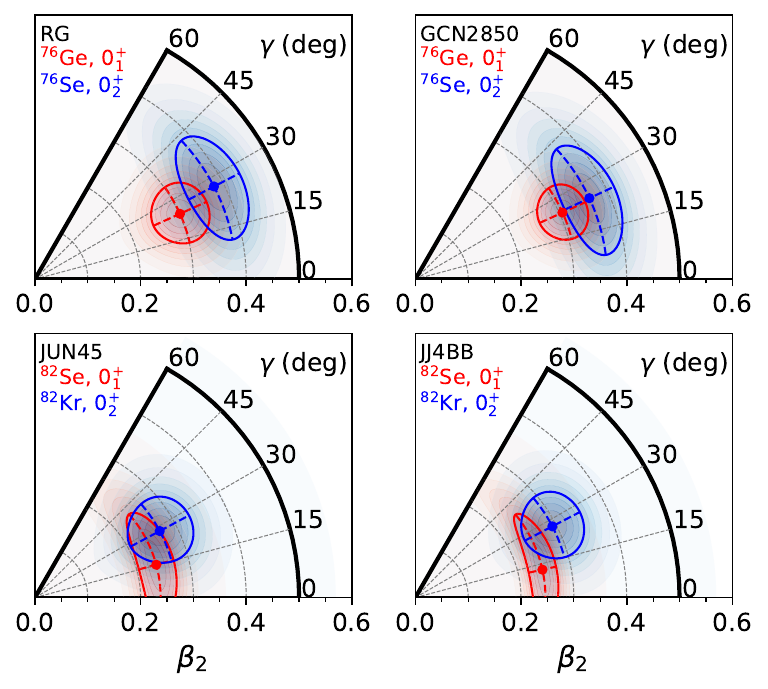}
    \caption{Deformation parameters $\beta_2,\gamma$ (circles) with 1$\sigma$ fluctuations  (ellipses) for $2\nu\beta\beta$ $0^+_{\rm gs}\xrightarrow[]{}0^+_{2}$ initial (red) and 
    final (blue) states. Top panels: results for the shell-model interactions giving the {longest} (RG) and {shortest} (GCN2850) $^{76}$Ge half-life. Bottom panels: results for the interactions giving shortest (JUN45) and longest (JJ4BB) $^{82}$Se half-life.  
    }
    \label{fig:76Ge_invariants}
\end{figure}

\begin{figure}[t]
 \centering
    \includegraphics[width=1.\linewidth]{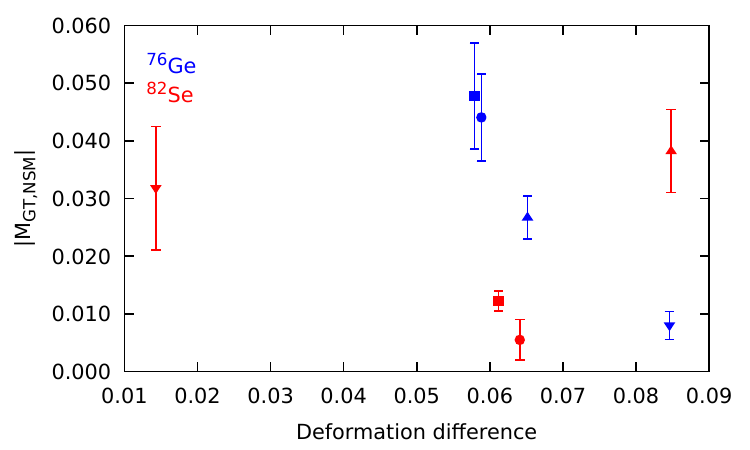}
    \caption{Leading-order $2\nu\beta\beta$ NSM NMEs as a function of the difference of deformation ($\delta_\text{def}$) for the $0^+_{\rm gs}\rightarrow0^+_{2}$ decays of $^{76}$Ge (blue) and $^{82}$Se (red). Results for the GCN2850 (squares), JUN45 (circles), JJ4BB (up triangles), and RG (down triangles) shell-model Hamiltonians. \textcolor{black}{The error bars cover all our results with bare and effective GT operators.}}
    \label{fig:76Ge_correlation}
\end{figure}

\subsection{$2\nu\beta\beta$, deformation and nuclear structure}
\label{sec:def}

We investigate this uncertainty by analyzing the nuclear structure of the initial and final states within each  nuclear Hamiltonian, using isoscalar effective charges $e_{p}=1.8e$ and $e_{n}=0.8e$ for protons and neutrons, respectively~\cite{ayangeakaa2023triaxiality}. In order to do so, we compute shape invariants, described in Sec.~\ref{sec:NSM}. Figure~\ref{fig:76Ge_invariants} shows the mean $(\beta_2,\gamma)$ values for each state, along with ellipses representing 1$\sigma$ deviations assuming a Gaussian bivariate distribution. For $^{76}$Ge, all interactions---also the ones not shown in the upper panels of Fig.~\ref{fig:76Ge_invariants}---suggest a rather rigid triaxial deformation ($\gamma$ covers approximately 15$\degree$), in contrast to the $\gamma$-soft $^{76}$Se $0^+_2$ state (with a spread of about 30$\degree$). We find that for GCN2850, JUN45, and RG the $^{76}$Ge ground state is similarly deformed (red circles in the upper panels in Fig.~\ref{fig:76Ge_invariants}), while for JJ4BB it takes a more deformed structure. There are also differences in the $^{76}$Se $0^+_2$  states, as evidenced by the blue circles in Fig.~\ref{fig:76Ge_invariants} upper panels. 

To illuminate the relation between nuclear structure and $2\nu\beta\beta$ decay, Fig. \ref{fig:76Ge_correlation} compares, for all interactions used for $^{76}$Ge and $^{82}$Se, the leading-order NMEs with the deformation difference, including triaxiality, between the initial and final states of the $2\nu\beta\beta$ $0^+_{\rm gs}\xrightarrow[]{}0^+_{2}$ decay,
\begin{equation}
    \delta_\text{def}=\sqrt{\beta_{2,i}^2+\beta_{2,f}^2-2\beta_{2,i}\beta_{2,f}\text{cos}(\gamma_f-\gamma_i)}\,,
\end{equation}
which is the distance between the central $(\beta_{2},\gamma)$ values. We observe a linear relation between the NMEs and deformation differences: more similar shapes correspond to larger NMEs, which can range over a factor of six. Notably, triaxiality plays a key role. For instance, in $^{76}$Ge, $\delta_\text{def}$ is similar for JUN45 and JJ4BB if $\gamma$ is neglected. This would break the linear correlation in Fig.~\ref{fig:76Ge_correlation}, as the corresponding NMEs differ by 50\% or so. Therefore, we conclude that a good part of the difference in $2\nu\beta\beta$ half-lives between Hamiltonians can be attributed to the structure of the initial and final states. 

Guided by this insight, we take a closer look at the excitation energies, $B(E2)$ transitions, and spectroscopic quadrupole moments,  $Q_s$, of $^{76}$Ge and $^{76}$Se low-lying states. While the energies and $B(E2)$s compare very well with experimental data~\cite{ayangeakaa2023triaxiality,Henderson:2019djd} for all interactions, Table~\ref{Table:Q_spectroscopic} indicates that only RG reproduces the $Q_s$ values in both nuclei. Using this criterion, the NSM $2\nu\beta\beta$ $0^+_{\rm gs}\rightarrow0^+_{\rm 2}$ half-life prediction would exceed the current limit by almost two orders of magnitude. 

Likewise, an analysis for $^{82}$Se and $^{82}$Kr reveals a preference for the JUN45 and JJ4BB Hamiltonians, for which the agreement with data \cite{Speidel:1998zz,Kavka:1995lcn,Lecomte:1977sza,Brussermann:1985zz} is good, but of somewhat less quality than for $A=76$. In contrast, RG misses important physics, such as the rotational behavior of $^{82}$Se. NSM half-lives diverge notably: while the JJ4BB prediction is {consistent} with the OBELIX $0^+_{\text{gs}} \rightarrow 0^+_2$ indication, the JUN45 one is {at least} an order of magnitude longer.

Indeed, Fig. \ref{fig:76Ge_correlation} shows that for this decay JJ4BB does not follow the linear relation between NMEs and $\delta_\text{def}$. This reflects that deformation, while important for $2\nu\beta\beta$ NMEs, is not the only relevant aspect. Along this line, we analyze the seniority structure of the initial and final states~\cite{menendez2011neutrinoless}. Our results reveal that the large JJ4BB NME for $^{82}$Se is {because the} cancellations that typically occur between different seniority components of the initial and final states~\cite{Caurier:2007wq} {do not occur when calculating this decay with this Hamiltonian}. Overall, our analysis does not favor either the shorter half-life prediction of JJ4BB, or the longer one of JUN45.




\begin{table}[t]
\caption{Spectroscopic quadrupole moments, $Q_s$, for low-lying states of $^{76}$Ge and $^{76}$Se, obtained with four shell-model interactions and compared with experimental values~\cite{NNDC}.}
\centering
\begin{tabular}{
    c
    @{\hspace{4pt}}S[table-format=-2.1(1)]
    @{\hspace{0pt}}S[table-format=-2.1, round-mode=figures, round-precision=2]
    @{\hspace{6pt}}S[table-format=-2.1, round-mode=figures, round-precision=2]
    @{\hspace{6pt}}S[table-format=-2.1, round-mode=figures, round-precision=2]
    @{\hspace{6pt}}S[table-format=-2.1, round-mode=figures, round-precision=2]
}
\hline \hline
 & \multicolumn{5}{c}{$Q_s$ ($e$fm$^2$)} \\
\hline
$^{76}$Ge & {Exp} & {JJ4BB} & {JUN45} & {GCN} & {RG} \\
\hline 
$2_1^+$ & -18(2) & -19.0 & 3.0   & -4.5  & -18.0 \\
$2_2^+$ &  20(3) &  20.0 & -0.7  & 4.4   & 19.0 \\
$4_1^+$ & -20(5) & -18.0 & -0.8  & 9.0   & -17.0 \\
\hline \hline
$^{76}$Se & {Exp} & {JJ4BB} & {JUN45} & {GCN} & {RG} \\
\hline
$2_1^+$ & -35(4) & 7.9   & 49.0  & 51.0  & -33.0 \\
$2_2^+$ &  19(4) & -9.0  & -42.0 & -46.0 & 31.0 \\
$4_1^+$ & -29(4) & 27.0  & 27.0  & 66.0  & -34.0 \\
\hline \hline
\end{tabular}
\label{Table:Q_spectroscopic}
\end{table}

\subsection{Lepton expansion and NLO contributions}


Table \ref{tab:results_tab} lists the relative contributions of the Taylor expansion in the lepton energies, $\varepsilon_\text{Taylor}$. {A negative sign denotes a longer $T^{2\nu}_{1/2}$, whereas positive contributions make it shorter}. These terms generally contribute about few percent to the $2\nu\beta\beta$ $0^+_{\rm gs}\to0^+_2$ half-life, consistently with the relative contributions found in QRPA calculations~\cite{Simkovic18}. However, for $^{82}$Se and $^{136}$Xe they become more relevant. This is because the leading-order NME undergoes a significant cancellation---already pointed out for $^{136}$Xe in Ref.~\cite{Jokiniemi:2022yfr}---which is absent in the subleading NMEs. The effect is most pronounced in $^{82}$Se due to its larger $\mathcal{Q}$. Also, note that the sign is different between the two interactions used for $^{136}$Xe, which is again related to the strong cancellations in the leading-order NME. 

NLO corrections for $2\nu\beta\beta$ $0^+_{\rm gs}\to0^+_2$ decays are also given in Table \ref{tab:results_tab} as $\varepsilon_\text{NLO}$. They are generally small, around 2\%, which is consistent with the expectation from $\chi$EFT~\cite{morabit20242nubetabetaspectrumchiraleffective}.
However, notable effects arise in $^{124}$Sn and $^{136}$Xe, due to very small leading-order NMEs combined with relatively large one-pion exchange NMEs. In all, this can lead to enhanced contributions, $|\varepsilon_{\rm NLO}| \sim(10 - {30})\%$ {for the bare GT operator. Interestingly, since cancellations in the leading GT matrix element are strongest with GT$^\text{eff}$ for the QX interaction, the NLO contributions can be comparable to the leading ones ($^{124}$Sn) or even dominant ($^{136}$Xe)}.

{Nonetheless, in general} these NLO contributions are very similar to the ones observed for $0^+_{\rm gs}\to0^+_{\rm gs}$ decays, listed in the last columns  {\ref{AnnexB}}. For $^{76}$Ge and $^{136}$Xe they have already been discussed in Ref.~\cite{morabit20242nubetabetaspectrumchiraleffective} for the GCN interactions. For these decays, NLO terms also contribute around 2\%, but they can be about twice as much for heavier nuclei. In fact, they reach up to 15\% for $^{136}$Xe, which features the smallest leading-order $2\nu\beta\beta$ NME.



\section{Summary}
In summary, we calculate NSM $2\nu\beta\beta$ half-lives for $0^+_{\rm gs} \rightarrow 0^+_2$ decays in $^{48}$Ca, $^{76}$Ge, $^{82}$Se, $^{124}$Sn, $^{130}$Te, and $^{136}$Xe, as well as the $^{124}$Sn $0^+_{\rm gs} \rightarrow 0^+_{\rm gs}$ decay. We expand the lepton energy denominator keeping the first three terms, and include NLO contributions given by $\chi$EFT. These terms typically remain below 5\%, but cancellations in the leading-order NME can enhance them, {so that they can become sizeable or even dominant.
We have used the bare GT operator, complemented with a phenomenological quenching, as well as a renormalized GT operator, both with and without quenching. In most cases, the two approaches yield relatively similar results, with an uncertainty dominated by the nuclear Hamiltonian used.}

Our predicted $0^+_{\rm gs}\xrightarrow[]{}0^+_2$ half-lives are, at least, two orders of magnitude longer than experimental lower bounds, except for two nuclei. For $^{76}$Ge, they are close to the current limit, but the NSM range, $({2-180})\cdot10^{24}$~yr, covers two orders of magnitude, depending on the nuclear Hamiltonian used. An analysis including triaxiality reveals that this is due to deformation difference between the initial and final nuclei. Even though all Hamiltonians describe well the spectroscopy of the relevant nuclei, the interaction that better reproduces spectroscopic quadrupole moments predicts the longest NSM half-life. Likewise, for $^{82}$Se the NSM range, $({0.4}-{190})\cdot10^{23}$~yr, is {consistent with} the recent measurement indication. A nuclear structure analysis does not show preference for longer or shorter predicted half-lives, but points to the importance of the seniority structure of the initial and final states in addition to their deformation. 

In future work we will aim to include short-range NLO NMEs, even though their coupling is currently unknown~\cite{morabit20242nubetabetaspectrumchiraleffective}, {as well as two-nucleon currents}. {We also aim to} further explore  the relation between nuclear structure, deformation, and $2\nu\beta\beta$ NMEs. A confirmation of the $^{82}$Se $0^+_{\rm gs}\xrightarrow[]{}0^+_2$ indication~\cite{Barabash:2025bxa}, and a possible measurement in $^{76}$Ge would be very valuable tests of our calculations. They would also help to reduce NME uncertainties in neutrinoless $\beta\beta$ decay.


\section*{Acknowledgements}
We would like to thank J. M. Fernández Varea for very insightful discussions as well as C. Brase for providing us the data of her previous calculations. This work is financially supported by MCIN/AEI/10.13039/501100011033 from the following grants: PID2023-147112NB-C22, CNS2022-135716 funded by the “European Union NextGenerationEU/PRTR”, PREP2023-002179 Doctoral Grant cofunded by FSE+ 2021-2027 and CEX2024-001451-M to the “Unit of Excellence María de Maeztu 2025-2031” award to the Institute of Cosmos Sciences; and by the Generalitat de Catalunya, through grant 2021SGR01095 and the predoctoral program AGAUR-FI ajuts Joan Oró 2025 FI (2025 FI-1 01213)  de la Secretaria d’Universitats i Recerca cofunded by FSE+ 2021-2027.

\clearpage

\onecolumn

\appendix
\section{}
\addcontentsline{toc}{section}{Anexos}\label{Annex}
\begin{table*}[h]
    \centering
    \caption{Previous predictions of the $0^+_{\rm gs}\xrightarrow[]{}0^+_2$ half-life (in yr) as well as of the $^{124}$Sn $0^+_{\rm gs}\xrightarrow[]{}0^+_{\rm gs}$ decay for different many-body methods: NSM, EFT$_{\beta}$, QRPA, and IBM.}
    \begin{tabular}{ccccc}
    \hline\hline
       Nuclei  &  NSM & EFT$_\beta$ & QRPA  & IBM  \\
    \hline
       \multirow{2}{*}{$^{48}$Ca  }& $(1.6-2.0)\cdot 10^{24}$\cite{PhysRevC.87.014320} & {$(1.57-52)\cdot10^{24}$}\cite{Jokiniemi:2022yfr} &   & $1.0\cdot10^{23}$\cite{PhysRevC.91.034304} \\
       & $2.7\cdot 10^{24}$\cite{Coraggio24} & $(0.62-49)\cdot 10^{24}$\cite{PhysRevC.98.045501} &  &   \\
       \hline
       \multirow{2}{*}{$^{76}$Ge}  & $(2.3-2.7)\cdot 10^{24}$\cite{gerda_collaboration_2_2015}& $(0.46-36)\cdot 10^{24}$\cite{Jokiniemi:2022yfr}  & $(1.2-5.8)\cdot 10^{23}$\cite{gerda_collaboration_2_2015} &$6.4\cdot 10^{24}$\cite{gerda_collaboration_2_2015}  \\
        & $5.3\cdot 10^{24}$\cite{Coraggio24}& $(0.28-170)\cdot 10^{24}$\cite{PhysRevC.98.045501} & $9.2\cdot 10^{24}$\cite{PhysRevC.91.034304}\\
         \hline
         \multirow{2}{*}{$^{82}$Se} &  $6.6\cdot 10^{24}$\cite{Coraggio24} &$(1.7-60)\cdot 10^{22}$\cite{Jokiniemi:2022yfr} &    &  $6.3\cdot 10^{23}$\cite{PhysRevC.91.034304}  \\
         & & $(0.21-39)\cdot10^{22}$\cite{PhysRevC.98.045501} &   &   \\
         
         \hline \noalign{\vskip 2pt}
         \multirow{1}{*}{$^{124}$Sn$(0^+_{\rm gs})$} & $1.6\cdot 10^{21}$\cite{horoi_shell_2016} & & $(2.8-3.4)\cdot 10^{21}$\cite{PhysRevC.91.054309} & $1.6\cdot 10^{21}$\cite{PhysRevC.91.034304} \\[2pt]
         
          \hline \noalign{\vskip 2pt}
       \multirow{1}{*}{$^{124}$Sn}  & $6.2\cdot10^{26}$\cite{horoi_shell_2016} & & $(8.2-10)\cdot 10^{24}$\cite{PhysRevC.91.054309}  & $6.3\cdot10^{25}$\cite{PhysRevC.91.034304} \\[2pt]
      
       \hline
       \multirow{2}{*}{$^{130}$Te}  & & $(0.78-33)\cdot 10^{25}$\cite{Jokiniemi:2022yfr} & $(0.70-1.7)\cdot 10^{24}$\cite{PhysRevC.91.054309}& $1.4\cdot 10^{25}$\cite{PhysRevC.91.034304}  \\
       & & $(0.25-63)\cdot 10^{25}$\cite{PhysRevC.98.045501}&  &   \\
       
       \hline
      \multirow{2}{*}{$^{136}$Xe}   & $(2.5-6.7)\cdot 10^{26}$\cite{Jokiniemi:2022yfr} & $(0.62-16)\cdot 10^{25}$\cite{Jokiniemi:2022yfr} & $(0.20-13)\cdot 10^{23}$\cite{Jokiniemi:2022yfr}  & $(1.5-3.7)\cdot 10^{25}$\cite{Jokiniemi:2022yfr} \\
      &  & $(0.23-3.2)\cdot 10^{25}$\cite{PhysRevC.98.045501} & $(1.3-6.9)\cdot 10^{23}$\cite{PhysRevC.91.054309} &  $2.5\cdot 10^{25}$\cite{PhysRevC.91.034304} \\

    \hline\hline
    \end{tabular}
    
    

    \label{tab:my_label}
\end{table*}
{
\section{}
\addcontentsline{toc}{section}{Anexos}\label{AnnexB}
\begin{table*}[h!]
    \centering
    \caption{Quenching factors for all nuclei (first column) and shell-model interactions (second column) used, as well as NLO contributions for $0^+_{\rm gs}\xrightarrow[]{}0^+_{\rm gs}$ decays for bare (sixth column) and effective (seventh column) GT operators. The third column lists the quenching values obtained from fits to GT decays, while the fourth (for the bare GT operator) and fifth (for the renormalized GT) columns correspond to those fitted to $T^{2\nu}_{1/2}(0^+_{\rm gs}\xrightarrow[]{}0^+_{\rm gs})$. The upper bound of $^{124}$Sn $q_{\beta\beta}$ is obtained from estimates inferred from studies of double electron capture in $^{124}$Xe~\cite{CoelloPerez:2018ghg,Nitescu:2024ppf}.}
\begin{tabular}{
    c            
    l            
    c            
    c  
    c     
    S[table-format=2.1(2)]  
    S[table-format=2.1(2)]  
}
\hline\hline\noalign{\vskip 3pt}
Nuclei & Interaction & $q_{\beta}$ \cite{CAURIER201262}  &  \multicolumn{2}{c}{$q_{\beta\beta}$} & \multicolumn{2}{c}{$\varepsilon_{\rm NLO,gs\xrightarrow[]{}gs}$ (\%)} \\
\hline\noalign{\vskip 3pt}
& & \text{GT$^{\rm bare}$} & \text{GT$^{\rm bare}$} & \text{GT$^{\rm eff}$} & \text{GT$^{\rm bare}$} & \text{GT$^{\rm eff}$}\\
\hline 
\multirow{2}{*}{$^{48}$Ca } & KB3G & 0.74 & 0.68  &  0.65 & 2.6(4) & 2.0(10) \\
& GXPF1A  & 0.74 & 0.64   & 0.63 & 2.3(6) & 1.9(11)\\
\hline 
\multirow{4}{*}{$^{76}$Ge }
& GCN2850   & 0.60 & 0.63  &  1.16  & 2.1(2) & 2.4(5)\\
 & JUN45     & 0.60 & 0.63  &  1.14 & 2.1(1) & 2.4(3)\\
& JJ4BB     & {}      & 0.64  & 1.15 & 2.1(1) & 2.5(5) \\
& RG & {}     & 0.55    &  1.02 & 2.0(2) & 2.1(3) \\
\hline 
\multirow{4}{*}{$^{82}$Se }
& GCN2850   & 0.60 & 0.53     & 0.97 & 2.0(4) & 2.2(2) \\
& JUN45     & 0.60 & 0.54      &  0.98 & 2.1(2) & 2.4(1)  \\
 & JJ4BB     & {}     & 0.58   &  1.21& 2.4(1) & 3.1(8)\\
& RG & {}     & 0.43      &  0.80 & 2.2(4) & 1.8(7) \\
\hline 
\multirow{2}{*}{$^{124}$Sn }
& GCN5082   & $0.57$ &  $0.42-0.50$    & $0.72-0.85$& 5.9(20)& 6.1(22)  \\
 & QX    &  & $0.67-0.82$     & $1.24-1.36$&  4.4(17)& 5.8(27)\\
\hline 
\multirow{2}{*}{$^{130}$Te }
& GCN5082   & 0.57 & 0.46      &  0.85 & 5.5(13) & 5.4(11) \\
 & QX    & {}  & 0.76       & 1.36 & 7.3(13) & 7.6(38) \\
\hline 
\multirow{2}{*}{$^{136}$Xe }
& GCN5082   & $0.45-0.57$ & 0.42      & 0.72 & 6.2(22) & 6.5(23) \\
 & QX    & {}   & 0.67      &  1.24 & 11.3(55) & 9.1(34)\\
\hline\hline
\end{tabular} \label{tab:AnnexB}
\end{table*}}

\clearpage

\twocolumn
\bibliography{apssamp}

\end{document}